\documentclass[acmsmall]{acmart}
\makeatletter
\def\@copyrightspace{\relax}
\makeatother

\renewcommand\footnotetextcopyrightpermission[1]{} 
\makeatletter
\renewcommand\@formatdoi[1]{\ignorespaces}
\makeatother

\usepackage{multirow}
\usepackage{booktabs}
\usepackage{subfig}
\usepackage{graphicx}

\begin{document}
%

\title[Interpretable Stochastic Block Influence Model]{Interpretable Stochastic Block Influence Model: measuring  social influence among homophilous communities}



%
%

\author{Yan Leng}
\affiliation{%
  \institution{MIT Media Lab, McCombs School of Business at UT-Austin}}
\email{yleng@mit.edu}
\author{Tara Sowrirajan}
\affiliation{%
  \institution{Harvard University, MIT Media Lab}}
\email{stara@mit.edu}
\author{Alex Pentland}
\affiliation{%
  \institution{MIT Media Lab}}
\email{pentland@mit.edu}

\begin{abstract}
Decision-making on networks can be explained by both homophily and social influence. 
While homophily drives the formation of communities with similar characteristics, social influence occurs both within and between communities. 
Social influence can be reasoned through role theory, which indicates that the influences among individuals depend on their roles and the behavior of interest. To operationalize these social science theories, we empirically identify the homophilous communities and use the community structures to capture the ``roles'', which affect the particular decision-making processes. 
We propose a generative model named Stochastic Block Influence Model and jointly analyze both the network formation and the behavioral influence within and between different empirically-identified communities. 
To evaluate the performance and demonstrate the interpretability of our method, we study the adoption decisions of microfinance in an Indian village. 
We show that although individuals tend to form links within communities, there are strong positive and negative social influences between communities, supporting the weak tie theory. 
Moreover, we find that communities with shared characteristics are associated with positive influence. 
In contrast, the communities with a lack of overlap are associated with negative influence. 
Our framework facilitates the quantification of the influences underlying decision communities and is thus a useful tool for driving information diffusion, viral marketing, and technology adoptions.

\end{abstract}


\maketitle


\section{Introduction} 
\noindent We are living in an increasingly connected society \cite{travers1967small,backstrom2012four,leng2018familiar,leng2018learning}. The connections among individuals foster information diffusion and enable the inter-dependencies in decision-making among peers. Therefore, understanding and modeling how hidden social influence changes individuals' decision-making are essential and critical for many practical applications, such as viral marketing, political campaigns, and large-scale health behavioral change \cite{fowler2008dynamic,pan2012decoding,leng2018rippling,leng2018contextual}. 

Homophily, the tendency of similar individuals to associate together, widely exhibits in various types of social networks, and governs the outcomes of many critical network-based phenomena \cite{mcpherson2001birds, kossinets2009origins, currarini2010identifying}. Salient features for homophily come from a wide range of sources, including age, race, social class, occupational, and gender \cite{mcpherson2001birds}. The complex nature of social relationships and high-dimensional characteristics of individuals thus determine the multi-dimensionality of homophily \cite{block2014multidimensional}. Homophily results in locally clustered communities and may affect network dynamics, such as information diffusion and product adoption. The Block Model has been applied to low-dimensional, pre-defined homophilous features and provides a building block to uncover underlying community structures\footnote{In this paper, we use community and  block interchangeably.} with high-dimensional homophily empirically \cite{abbe2017community}.

Social influence is widely studied in economics and computer science literature due to its importance in understanding human behavior. In economics, researchers focus on causally disentangling social influence from homophily with randomization strategies, such as propensity score matching \cite{aral2009distinguishing}, behavioral matching \cite{leng2018rippling} and regression adjustment \cite{angrist2014perils}. In the computer science literature, researchers focus on maximizing the likelihood of the diffusion path of influence by proposing different generative processes \cite{gomez2012inferring,gomez2013structure, myers2012information, yang2010modeling}. 
These works focus on the strength or the pathways of social influence, and they do not link social influence to the underlying homophilous communities and the network formation process.

There exist two theories explaining how local communities affect information diffusion \cite{weng2013virality} and contagion in decision-making \cite{leng2018rippling, golub2012homophily}. On the one hand, homophily and the requirement of social reinforcement for behavioral adoption in complex contagion theory indicate that influence tends to be localized in homophilous communities \cite{mcpherson2001birds, centola2007complex}. In other words, behavioral diffusion and network formation are endogenous, explaining the phenomenon of within-community spreading \cite{pin2016stochastic, weng2013virality}. On the other hand, the weak ties theory \cite{granovetter1977strength} implies that bridging ties between communities facilitate the spreading of novel ideas. As empirical evidence, Ugander shows that reinforcement from the multiple communities, rather than from the same communities, predicts higher adoption rates \cite{ugander2012structural}. With these two competing theories, we seek to understand whether social influence spreads locally within each homophilous community or globally to other communities taking advantage of the long ties. 

{Role theory posits that ``the division of labor in society takes the form of interaction among heterogeneous specialized positions'' \cite{biddle1986recent}. 
That is to say, depending on the social roles and the behavior of interest, the underlying interactions and norms for decision-making are different. 
Motivated by this proposition, we aim to develop a method to associate social influence with the underlying communities, which are associated with the behavior of interest}. 
{To formalize this idea, we propose a generative model to understand how social influence impacts decision-making by inferring the spreading of influence across empirically-identified blocks. Our framework jointly uncovers the underlying blocks and infers two types of relationships across these blocks: social interaction and social influence. Different from the Stochastic Block Model, the observed individual decisions are used to inform the communities, as complementary to the observed network. 
Along with this, we infer an influence matrix as the social influence across different communities. 
This influence matrix reveals the hidden social influence at the community level, which would otherwise be impossible to observe and generalize. 

As a case study, we experiment on the diffusion of microfinance in an Indian village and perform extensive analysis on the influence matrix estimated from the model. We find that even though social relationships are denser within communities, social influence mainly spreads across communities.  
This may be explained by the importance of cross-community weak ties \cite{granovetter1977strength} and the strength of structural diversity \cite{ugander2012structural}. Our generative framework and subsequent understanding of how social influence operates are informative for practical applications, such as viral marketing, political campaigns, and large-scale health-related behavioral change \cite{fowler2008dynamic,pan2012decoding,leng2018rippling}.
}

\paragraph*{Contributions} To summarize, the Stochastic Block Influence Model (SBIM) developed in our study makes the following contributions to the literature:
\begin{itemize}
  \item SBIM integrates networks, individual decisions, and characteristics into the generative process. It jointly infers two types of relationships among empirically-identified communities: social connection and social influence. Moreover, our model flexibly accommodates both positive and negative social influences. 
  \item {Our model is motivated by role theory, which posits that individuals make decisions depending on the context of the decision type \cite{biddle1986recent}, e.g., adopting microfinance as opposed to adopting healthy habits. To achieve this, we allow the underlying community to vary with the behavior of interest. }
  \item We perform a case study on the adoption of microfinance in an Indian village. Moreover, we demonstrate the interpretability of our model with a detailed analysis of the influence structure. 
  \item The analysis from our study can be used for designing network interventions and marketing strategies. For example, we show that communities with smaller overlaps in characteristics exert negative influences on one another. Therefore, marketing firms should encourage individuals to communicate with neighbors in the same community, such as inviting these individuals together to an informational event to promote the positive influence among them.
  \item SBIM bridges the rich Stochastic Block Model and the social contagion literature. It opens up future opportunities to adapt to other variations of SBM, e.g., degree-regularized SBM \cite{gao2018community} or SBM adjusted for power-law distributions \cite{qiao2018adapting}. 
\end{itemize}

The remaining sections are organized as follows. We describe the literature in Section~\ref{sec:related}. In Section~\ref{sec:model}, we introduce the proposed Stochastic Block Influence Model. Then, we test the method in Section~\ref{sec:experiment} and analyze the results on a real-world data set in Section~\ref{sec:result}. In Section~\ref{sec:conclusion}, we summarize the paper with practical applications and future work.

\section{Related literature}
\label{sec:related}
\paragraph*{Contagion models}
There are two prominent theories in the literature for explaining the propagation of social influence \cite{ugander2012structural,bond201261,aral2009distinguishing, leng2018rippling}, i.e., simple contagion and complex contagion. Simple contagion theory assumes that individuals will adopt the behavior as long as they have been exposed to the information \cite{granovetter1977strength}, which is a sensible model for epidemics and information spreading. Complex contagion theory, on the other hand, requires social reinforcement from neighbors to trigger the adoption \cite{centola2007complex}. Many studies have shown that complex contagion explains behaviors such as registration for health forums \cite{Centola1194}. 

These exposure-based models bear analytical simplicity, however, do not allow social influence to be negative, i.e., the adoption decision of one's neighbors might decrease, rather than increase, the likelihood of one's adoption decision. Moreover, they typically are not able to capture the heterogeneity of social influence \cite{leng2018contextual}. In this paper, we propose a model to account for negative and heterogeneous influence.  

\paragraph*{Stochastic Block Model}
The Stochastic Block Model is a statistical model for studying latent cluster structures in network data \cite{abbe2017community}. SBM generalizes the Erdos-Renyi random graph model with higher intra-cluster and lower inter-cluster probability. The traditional SBM only infers the community structures from network connections. However, when contextual information on nodes is available, leveraging information from different sources facilitates the inference. In recent statistics literature, there has been some interesting work on utilizing covariates to infer the block structures. For example, Binkiewicz et al. present a covariate-regularized community detection method to find highly connected communities with relatively homogeneous covariates \cite{binkiewicz2017covariate}. They balance the two objectives (i.e., the node covariance matrix and the regularized graph laplacian) with tuned hyper-parameters. Yan et al. propose a penalized optimization framework by adding a k-means type regularization \cite{yan2019covariate}. This framework enforces that the estimated communities are consistent with the latent membership in the covariate space. 

{Though these variations to SBM utilize auxiliary information on individual nodes, they specify the importance of recovering the network and the smoothness of covariates on the network, on an ad-hoc basis. 
{Different from these models, we take advantage of role theory \cite{biddle1986recent} and utilize the decision-making process on the network that could also inform community detection. 
For example, let us assume professional communities are more useful for the adoption of technologies at work, and social communities are more useful for the adoption of social apps. 
The underlying communities depend on the role and behavior of interest because social influence spreads through some specific network links in different applications. }}

\section{Methodology}
\label{sec:model}

\subsection{Stochastic Block Influence Model} 

\paragraph*{Notations} 
Assume a random graph $G(\mathcal{V}, \mathcal{E})$ with $N$ individuals in node set $\mathcal{V}$ and edge set $\mathcal{E}$. It is partitioned into $C$ disjoint blocks ($\mathcal{V}_1, ..., \mathcal{V}_C$), and the proportion of nodes in each block $c$ is $\pi_c$, and $\sum_{c=1}^C \pi_c = 1$. $\mathbf{A} \in \mathbb{R}^{N \times N}$ represents the adjacency matrix. $\mathbf{A}_{ij} = 1 $ if $i$ and $j$ are connected, and $\mathbf{A}_{ij} = 0$ otherwise. Let matrix $\mathbf{B} \in \mathbb{R}^{C\times C}$ denote the inter-block and intra-block connection probability matrix. Let $\mathbf{M}_i$ be the block assignment of individual $i$ and summing over C  blocks, we have  $\sum_{k=1}^C \mathbf{M}_{ik} = 1$. Together, we combine the  block vector of all individuals in the matrix $\mathbf{M} \in \mathbb{R}^{N \times C}$. Therefore, the probability of a link between $v_i$ and $v_j$ between two separate blocks $\mathcal{V}_k$ and $\mathcal{V}_l$ as $P \big((v_i, v_j) \in \mathcal{E} | v_i \in \mathcal{V}_k, v_j \in \mathcal{V}_l \big) = p_{ij}$. $\mathbf{y} \in \mathbb{R}^{N}$ is a binary vector representing individuals' adoption behaviors. Let $\mathbf{X}_i \in \mathbb{R}^D$ {represent demographic features}, where $D$ is the number of covariates. 
We use $\mathbf{F} \in \mathbb{R}^{C \times C }$ to represent the block-to-block influence matrix. {Finally, $\mathbf{h}$ is a binary vector, capturing whether or not each individual is aware of the product at the beginning of the observational period. 
For a new product, $\mathbf{h}$ is sparse, while for a mature product, $\mathbf{h}$ is dense. }

\paragraph*{Model formulation}
Extending SBM to utilize the network, adoption decisions, and sociodemographic features, we propose the Stochastic Block Influence Model, abbreviated as SBIM. Linking {the latent communities} to their sociodemographic composition, we reveal the underlying nature of high-dimensional homophily in a data-driven fashion rather than using pre-defined communities using observed sociodemographics, e.g., race or occupation. Solely using pre-defined homophilous characteristics does not aptly capture the multiplex characteristics that define individuals and their social ties. In other words, individuals are associated with different communities, each of which is formed by various homophilous characteristics. Neighbors belonging to different communities may influence the focal individuals differently. 

Let us illustrate this using the adoption of microfinance in an Indian village. 
It is reasonable to posit that several traits define the diverse nature of individuals - different professions, castes, education levels, and a variety of other demographic features. 
Let us take one particular individual, who is an educated worker of a lower caste, for example. This individual belongs with varying degrees of affiliation to different communities: perhaps most strongly affiliated to a group of a certain level of education and less strongly affiliated with another group of a majority of a lower caste. This mixed membership captures the realistic nature of our social relationships and characteristics. 
Within such a village with multi-dimensional homophily, how can we understand who influences this individual and what processes are involved in that individual's decision making? 
Specifically, she could be influenced both by neighbors belonging to different communities characterized by specific educational backgrounds, professions, and castes. 
The data-driven multi-dimensional block aspect of the model allows us to capture these critical, hidden relationships. 


Next, we formalize our model. To jointly infer how influence spreads within and across communities, we desire a model with the following properties: 
\begin{enumerate}
  \item The model leverages both the observed friendship network structure and the adoption behavior to infer the underlying communities. 
  \item The link formation and social influence between two individuals are jointly determined by their underlying communities. 
\end{enumerate}


For each individual pair $\{i, j \}$, depending on their community assignment vectors, the predicted link $\tilde{\mathbf{A}}_{ij}$ is generated according to the connection probability matrix, $\mathbf{B}$. In particular, the probability of the existence of a link between $i$ and $j$ is, 
\begin{equation}
  \mathbb{P} ( \tilde{\textbf{A}}_{ij} = 1 | \mathbf{M}, \mathbf{B}) = (\mathbf{M} \mathbf{B} \mathbf{M}^T)_{ij}. 
  \label{eq:link}
\end{equation}

\begin{figure}     
       \centering              
        \includegraphics[width=.6\linewidth]{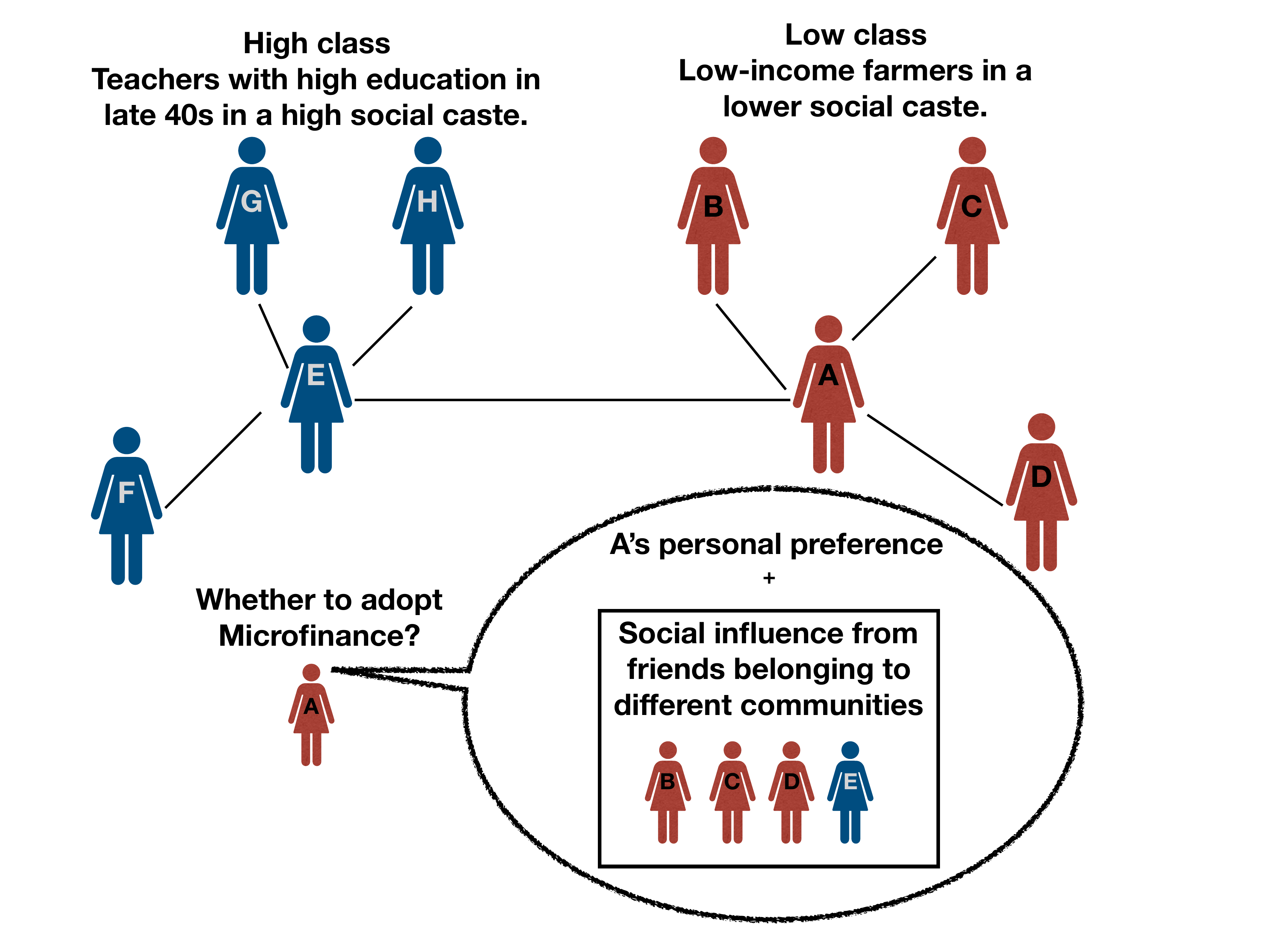}
         \caption{Graphical representation of the Stochastic Block Influence Model (SBIM). Assume there are two communities, a high socioeconomic status (SES) group (colored in dark blue) and low SES group (colored in dark red), characterized by multi-dimensional sociodemographic features. The two groups have higher intra-class connection probability and lower inter-class connection probabilities. The decision-making of A is jointly influenced by her preferences, as well as her neighbors from the same and different communities. }
         \label{graph_rep}
\end{figure}

Next, we discuss how our model incorporates individual characteristics and adoption decisions. The adoption likelihood depends on individuals' characteristics and on the influence of their neighbors who have already adopted \cite{katona2011network}. The generative model builds upon the communities a particular individual $i$, and $i$'s neighbors belong to, as well as the community-to-community matrix $\mathbf{F}_{ij}$. 
Each individual makes a decision on whether or not to adopt in order to maximize her utility. The utility of $i$ depends on her own preferences and the aggregated influence from neighbors.
The pairwise influence depends on the communities $i$ and her neighbors belong to. 
We illustrate how influence and communities affect one's decision-making in Figure~\ref{graph_rep}. Let us consider individual A, who has three friends, B, C, and D, belonging to a lower socioeconomic status (SES) group (as colored in red), and one friend, E, belonging to a higher SES group (as colored in blue). 
The adoption likelihood of A is a function of her own preferences as well as the influence from her friends B, C, D, and E. The strength of the influence depends on the corresponding communities of A and her friends (B, C, D, and E). 

{
More generally, the adoption likelihood of a user, $ \hat{\mathbf{y}}$, is defined as, 
\begin{equation}
 \hat{\mathbf{y}}_i = \text{logit} \big( \boldsymbol{\beta} \mathbf{X}_i + \sum_j \big( (\mathbf{M} \mathbf{F} \mathbf{M}^T )  \circ ( (\mathbf{h} \otimes \mathbf{1}) \circ \mathbf{A} ) \big)_{ji } + \epsilon_i \big), 
   \label{eq:decision}
\end{equation}

\noindent where $\circ$ is the element-wise matrix multiplication. The first term, $\boldsymbol{\beta} \mathbf{X}_i$, measures the adoption decision conditioned on $i$'s sociodemographic features if there were no social influence, where $\boldsymbol{\beta} \in \mathbb{R}^D$ and $D$ is the dimension of the covariates. The second term aggregates the influence of $i$'s neighbors. $\epsilon_i$ is the idiosyncratic error term. Without loss of generality, we assume $\epsilon_i \sim \mathcal{N}(0,1)$. 

For a mature product that everyone is aware of, we can simplify Equation~(\ref{eq:decision}) as, 

\begin{equation}
 \hat{\mathbf{y}}_i = \text{logit} \Big( \boldsymbol{\beta} \mathbf{X}_i + \sum_{j=1}^N \big( (\mathbf{M} \mathbf{F}  \mathbf{M}^T ) \circ \mathbf{A} \big)_{ji} + \epsilon_i  \Big). 
   \label{eq:decision2}
\end{equation}
}



Equation~(\ref{eq:decision}) only accounts for the influence among direct neighbors.
Note that in a small-scale network, {it is reasonable to assume that there does not exist higher-order social influence.} In a large-scale network, Leng et al. show that social influence spreads beyond immediate neighbors \cite{leng2018rippling}. For these applications, our model can be easily adapted to higher-order influence by summing up the powers of the adjacency matrix $A$ to account for multiple degrees of separation \cite{leng2018contextual}.



\subsection{Generative process}
For the full network, the model assumes the following generative process, which defines a joint probability distribution over $N$ individuals, based on node-wise membership matrix $\mathbf{M}$, block-to-block interaction matrix $\mathbf{B}$, block-to-block influence matrix $\mathbf{F}$, attributes' coefficients $\boldsymbol{\beta}$, observed friendship network $\mathbf{A}$, observed attributes $\mathbf{X}$, observed adoption decision $\mathbf{y}$.

\begin{enumerate}
\item For each node $v_i \in {\mathcal{V}}$, draw a $C$-dimensional mixed membership vector $\mathbf{M}_i \sim \text{Dirichlet}({c})$. 
\item For the connection probability from community $k$ to $l$ in the block-to-block connectivity matrix, draw $\mathbf{B}_{kl}$ $\sim$ $\text{Beta} (a, b)$. 
\item For the influence from community $k$ to $l$ in the block-to-block influence matrix, draw $\mathbf{F}_{kl}$ $\sim$ $\mathcal{N}$ $(\mu_F, \sigma_F)$. 
\item For each attribute in $\boldsymbol{\beta}$ indexed by $d$, draw the coefficient $\boldsymbol{\beta}_d \sim \mathcal{N}({\mu}_{b}$,${\sigma}_{b})$. 
\item Draw the connection between each pair of nodes $v_i$ and $v_j$, $\hat{\mathbf{A}}_{ij}$, according to Equation~(\ref{eq:link}). 
\item Draw the adoption decision  $\hat{\mathbf{y}}_{i}$, according to Equation~(\ref{eq:decision}). 

\end{enumerate}

For abbreviation, we denote $\mathcal{Z}$ as set of the hidden variables, $\mathcal{Z} = \{\mathbf{M}, \boldsymbol{\beta}, \mathbf{B}, \mathbf{F} \}$ and $\theta$ as the set of hyperparameters, where $\theta = \{c, a, b, \mu, \sigma, \mu_b, \sigma_b\}$. 

The posterior distribution defined by the generative model is a conditional distribution of the hidden block structure and relationships given the observed friendship network and adoption behavior, which decomposes the agents into $C$ overlapping blocks. The posterior will place a higher probability on configurations of the community membership that describe densely connected communities as well as stronger (positive or negative) influences. We present a visualization in Figure~\ref{fig:adj}, which illustrates that the posterior superimposes a block structure on the original network. The details of the data we use are described in Section~\ref{sec:experiment}. 

\paragraph*{Inference}
\label{infer}
The posterior of SBIM is intractable, similar to many hierarchical Bayesian models \cite{bayarri2003hierarchical}. Therefore, we use the Markov Chain Monte Carlo (MCMC) algorithm as an approximate statistical inference method to estimate the parameters. MCMC draws correlated samples that converge in distribution to the target distribution and are generally asymptotically unbiased. 

There are different MCMC methods, including Gibbs sampling, Metropolis-Hastings, Hamiltonian Monte Carlo, and No-U-Turn Sampler (NUTS). Gibbs sampling and Metropolis-Hastings methods converge slowly to the target distribution as they explore the parameter space by random walk \cite{2011arXiv1111.4246H}. HMC suppresses the random walk behaviors with an auxiliary variable that transforms the problem by sampling to a target distribution into simulating Hamiltonian dynamics. However, HMC requires the gradient of the log-posterior, which has a complicated structure in our model. Moreover, it requires a reasonable specification of the step size and a number of steps, which would otherwise result in a substantial drop in efficiency \cite{hoffman2014no}. 

Therefore, we apply NUTS, a variant to the HMC method, to eliminate the need for choosing the number of steps by automatically adapting the step size. Specifically, NUTS builds a set of candidate points that spans the target distribution recursively and automatically stops when it starts to double back and retrace its steps \cite{hoffman2014no}. We use the NUTS algorithm implemented in Python PyMC3 \cite{salvatier2016probabilistic}.

\section{Experiments}
\label{sec:experiment} 
\paragraph*{Data description}
We study the adoption of microfinance in an Indian village collected by the Abdul Latif Jameel Poverty Action Lab (J-PAL) \cite{banerjee2013diffusion}\footnote{The village we study is indexed by 64.}. {In 2007, a microfinance institution introduced a microfinance program to some selected Indian villages. In early 2011, they collected information about whether or not the villagers had adopted microfinance. Because the village is fairly small (257 villagers) and microfinance had been on the market for four years when JPAL collected individuals' adoption decisions, it is reasonable to assume that everyone in the village was aware of microfinance, which is hence a mature product. Therefore, we use Equation~(\ref{eq:decision2}) as the decision-making function.} The data contains information about self-reported relationships among households and other amenities, including village size, quality of access to electricity, quality of latrines, number of beds, number of rooms, the number of beds per capita, and the number of rooms per capita. {These types of demographic features} are used as the independent variables. The outcome variable is the adoption decision of = microfinance. The microfinance institution asked the villagers to self-report other villagers they considered as friends.

\paragraph*{Baseline}
We use the Random Forest with sociodemographics and the hidden community learned by spectral clustering on the adjacency matrix as the independent variables. In this way, we use the same information in SBIM and the baseline. Spectral clustering uses the second smallest eigenvector of the graph laplacian as the semi-optimal partition \cite{ng2002spectral}.

\paragraph*{Model training} To train our model and evaluate the performance for a particular $C$, the number of block, we cross-validated by randomly splitting the data into 75\% training samples and 25\% test samples. We repeat this process ten times. With NUTS, we obtain the point estimates for all latent variables in $\mathcal{Z}$\footnote{Some critical hyperparameters for NUTS are the number of burn-in samples, the number of samples after burn-in, the target acceptance probability, and the number of chains. For all of our NUTS sampling runs, we burn 3,000 samples to ensure that MCMC mostly converges to the actual posterior distribution. The number of samples after burn-in is 500; usually, only less than ten samples (among the 500) are diverging. Next, we select the target acceptance probability to be 0.8. At the end of each run, we average across the 500 samples to derive point estimates for all latent variables.}. We then re-run our model (as previously described) with all latent variables fixed to the estimates on the test dataset. This step returns the predicted adoption probability for each villager in the test data. 

To choose the optimal number of block, we first tune the model for $C \in \{2,6,10,14\}$ and then calculate the average loss. We observe a negative parabolic trend with the loss peaking at its lowest at $C = 10$ blocks, so we use this optimal number of block for further evaluation.



\paragraph*{Model evaluation}
Since the dependent variable in our data is imbalanced, we evaluate our method using the AUC, which is the area under the Receiver-Operating-Characteristics curve plotted by the false positive rate and correct positive rate for different thresholds. We define a loss metric during the training period to select the best configurations. It is formulated by the negative of the standard improvement measure, which is the absolute improvement in performance normalized by the room for improvement. This measure captures the improvement of our method compared to the baseline. Since we have a small test set, a randomly-drawn test set may be harder to predict than others. Measuring the relative improvement ensures that the composition of the test set does not bias the performance due to sample variation. This metric is formulated by 
\begin{equation}
L = \frac{\text{Baseline}_{\text{ test AUC}} - \text{SBIM}_{\text{ test AUC}}}{1 - \text{Baseline}_{\text{ test AUC}}},
\label{eq:loss}
\end{equation}
\noindent where the AUC of the baseline and SBIM on the test split in cross-validation are represented as $\text{Baseline}_{\text{ test AUC}}$ and $\text{SBIM}_{\text{ test AUC}}$, respectively.



Our model has seven hyperparameters, $\theta = \{c, a, b, \mu, \sigma, \mu_b, \sigma_b\}$\footnote{The ranges from which these hyperparameters were sampled are as follows: $\mu_{b} \in \left[-2, 2\right] $, $\sigma_{b} \in \left[-0.1, 1\right]$, ${c} \in \left[0.5, 1.5\right] $, $\mu_{{F}} \in \left[-6, 6\right] $, and $\sigma_{{F}} \in \left[0.1, 3\right] $. We let $ a, b = 2$ for a reasonable and non-skewed prior. }. 
Since the parameter space is large, we adapt a bandit-based approach to tune the parameters developed called Hyperband \cite{li2017hyperband}. The Hyperband algorithm adaptively searches for configurations and speeds up the process by adaptive resource allocation and early-stopping. Our adaptation of this algorithm allows each configuration tested to run with full resources due to the sampling procedure used in our methodology, allowing NUTS to run consistently across all configurations.

\paragraph*{Performance}
We compare the performance of our model with the baseline in Table \ref{tab:performance}. We observe that our method outperforms random forest in the test set by 13.8\% by the improvement metric in Equation~(\ref{eq:loss}). Both models overfit the training set and the baseline overfit comparatively more. 

\begin{table}[h]
\caption{Model and baseline performance} 
\label{tab:performance } 
\begin{center}
\begin{tabular}{l |c| c} 
\hline
& Mean & Standard deviation \\ 
\hline 
Baseline train AUC &0.901 & 0.010 \\ \hline
SBIM train AUC & 0.805 & 0.022 \\\hline
Baseline test AUC & 0.610 & 0.095 \\ \hline
SBIM test AUC & 0.664 & 0.062 \\ 
\hline 
\end{tabular}
\label{tab:performance} 
\end{center}
\end{table}

\section*{Analysis and discussions}
\label{sec:result}
\begin{figure}[h]
\centering
\includegraphics[width=.6\linewidth]{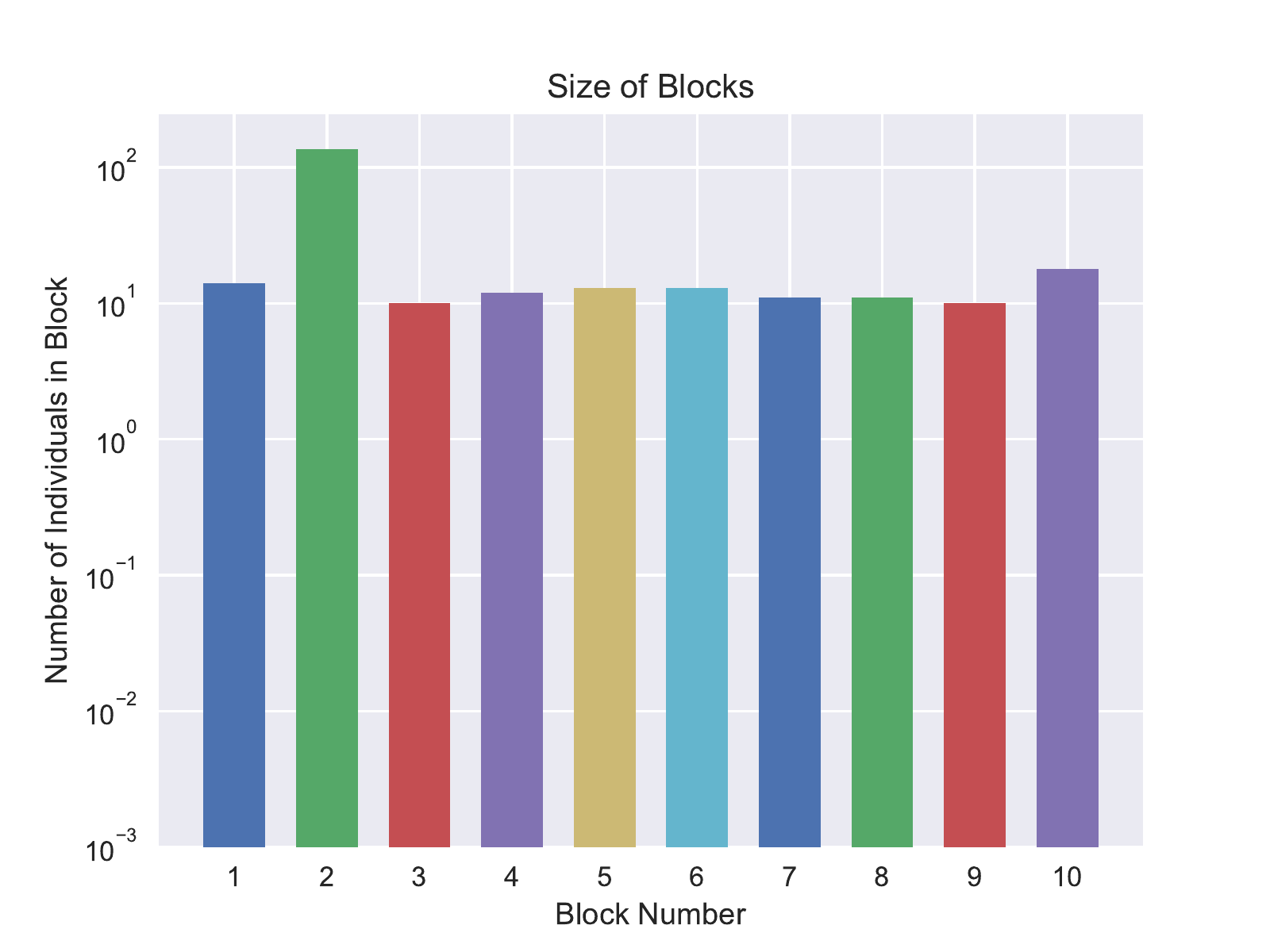}
\caption{Size of each social block. The y-axis corresponds to the number of individuals in the block, and the x-axis is the corresponding block index.}
\label{fig:size}
\end{figure}

\begin{figure}[h!t]
\centering
 {\includegraphics[width=0.6\linewidth]{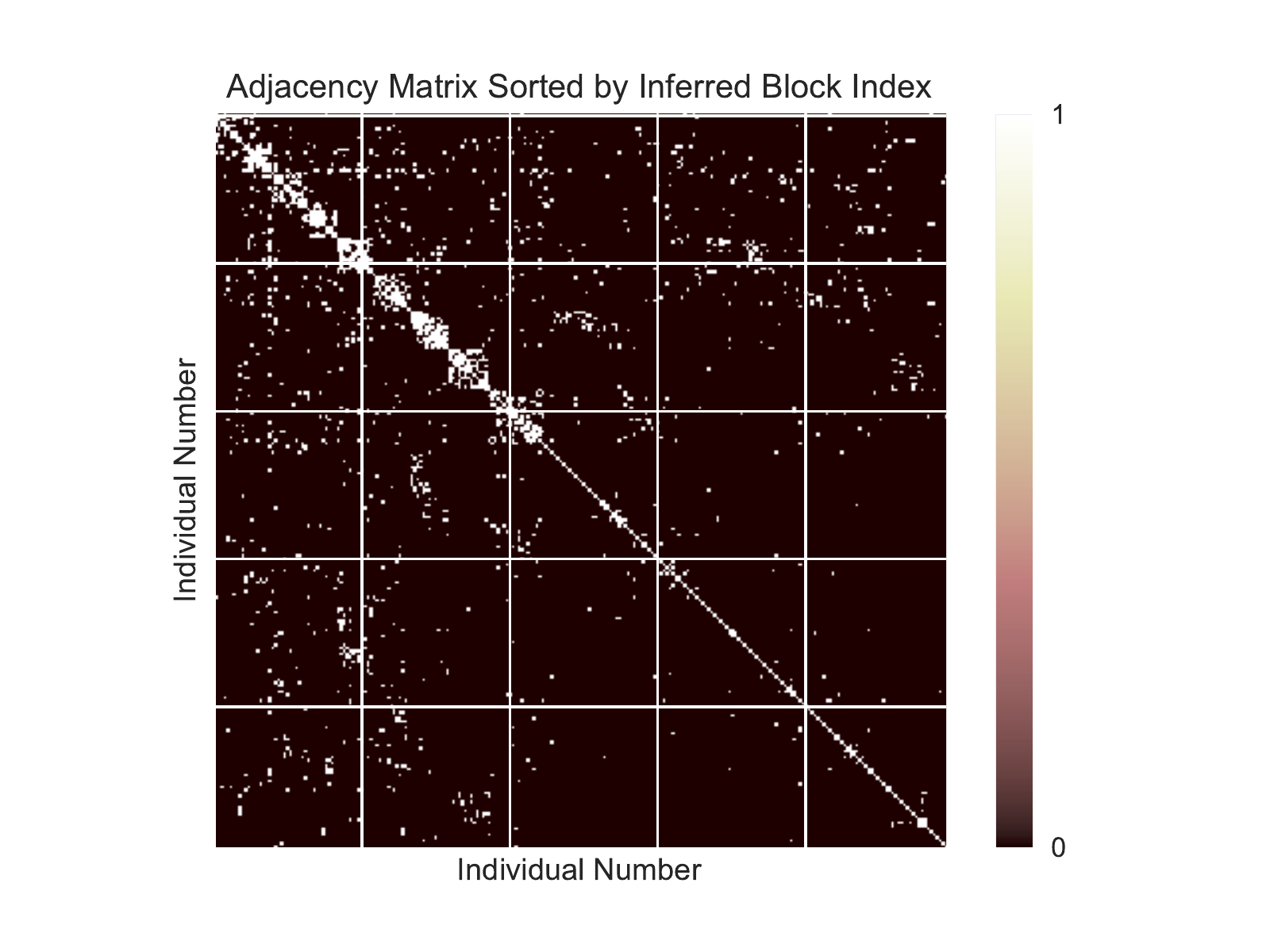}}
 \caption{Adjacency matrix sorted by the inferred block index. The x-axis and y-axis correspond to the indices of individuals. The white and black cells correspond to the existence and the non-existence of edges. We can clearly observe the underlying communities from the network.}
 \label{fig:adj}
\end{figure}

\paragraph*{Size of communities and interaction matrix}
We present the size of each social block in Figure~\ref{fig:size}. Social block two is larger than the other blocks, and the sizes of the rest are similar. 
{This aligns with our intuition that many individuals belong to a majority group while several niches, minority communities also exist.}
We represent the adjacency matrix sorted by this inferred block index from smallest to largest block in Figure~\ref{fig:adj}.  We see that there are many links within all of the blocks along the diagonal, demonstrating that the block model is meaningful and captures more links within than across blocks. The largest block, furthest along the diagonal, is comparatively sparser. 

\paragraph*{Block type}

We can associate individuals' sociodemographic characteristics with the individuals who belong to each block to generalize block type as consisting of characteristics such as high or low SES, homogeneous or diverse, and skilled or less educated, as depicted in Table \ref{blockChar}. In this example, each block is associated with a qualitative type, and the attributes within that block leading to such characterizations are described. Lower or higher SES blocks are designated by caste composition, education levels, and profession types. Homogeneous or diverse blocks are designated by some professional composition, caste types, mother tongue language composition, gender imbalance, and what fraction of village inhabitants are natives.

\begin{figure*}[ht!]
\centering 
   \includegraphics[width=.32\linewidth]{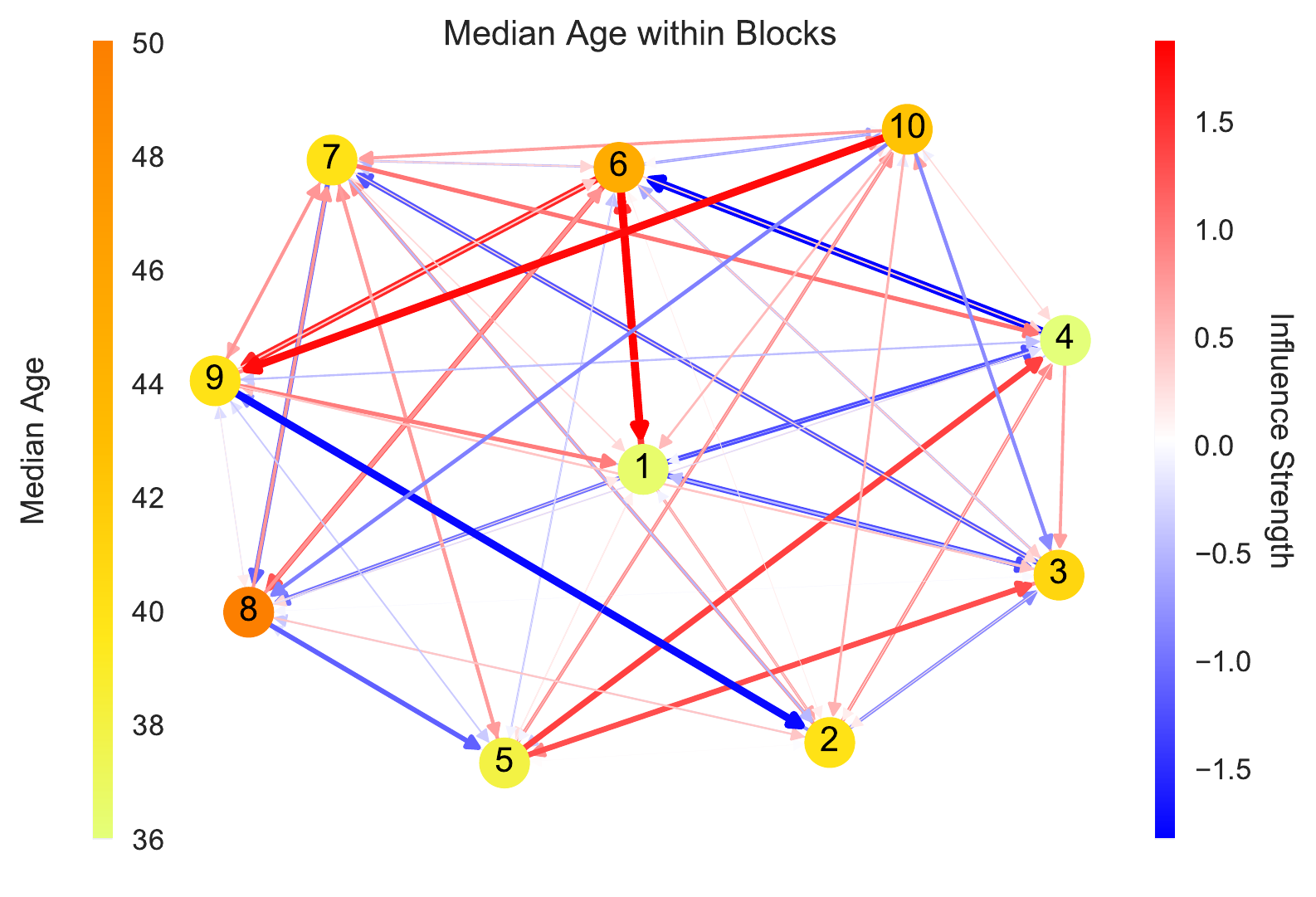} 
   \includegraphics[width=.32\linewidth]{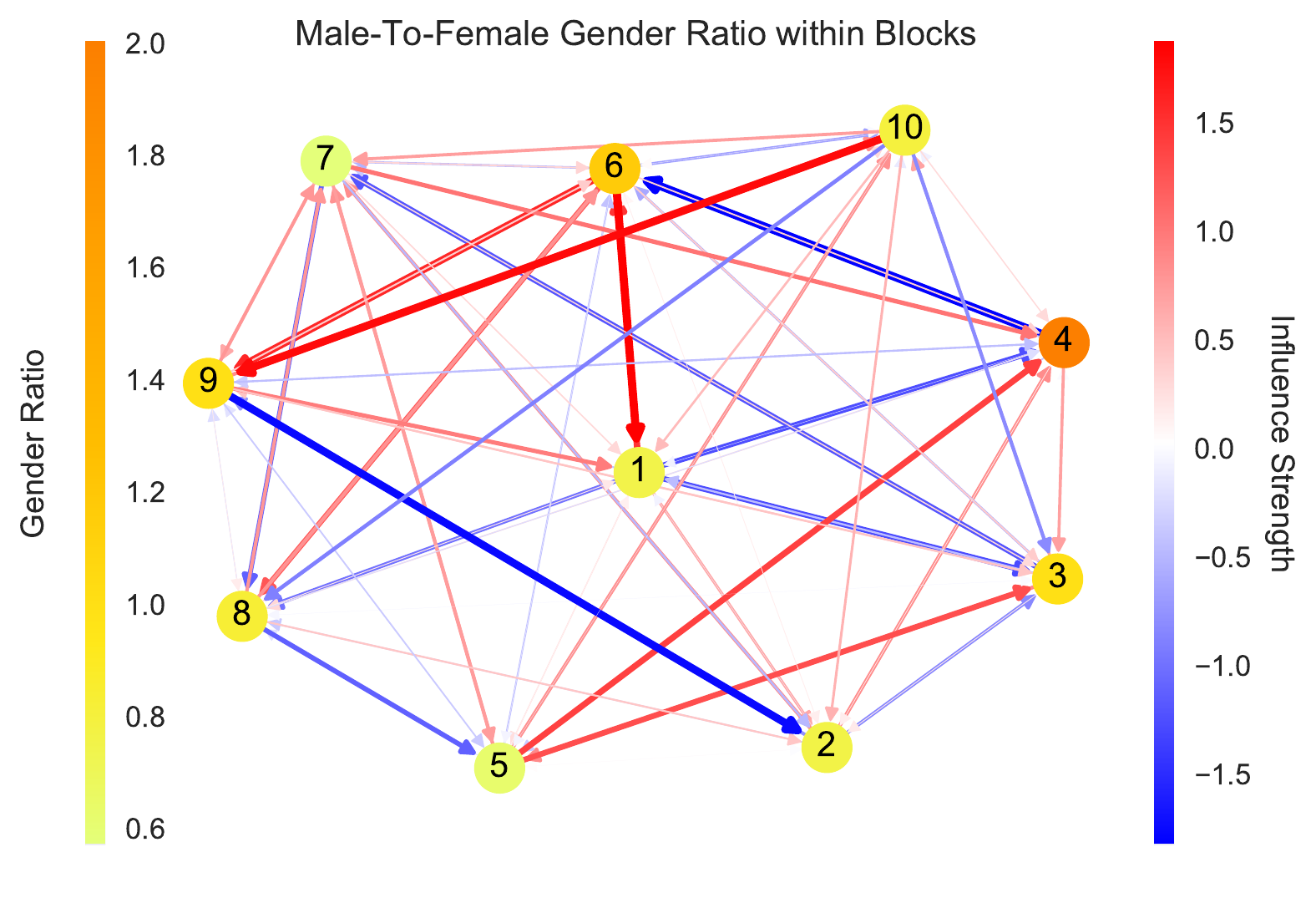}
   \includegraphics[width=.32\linewidth]{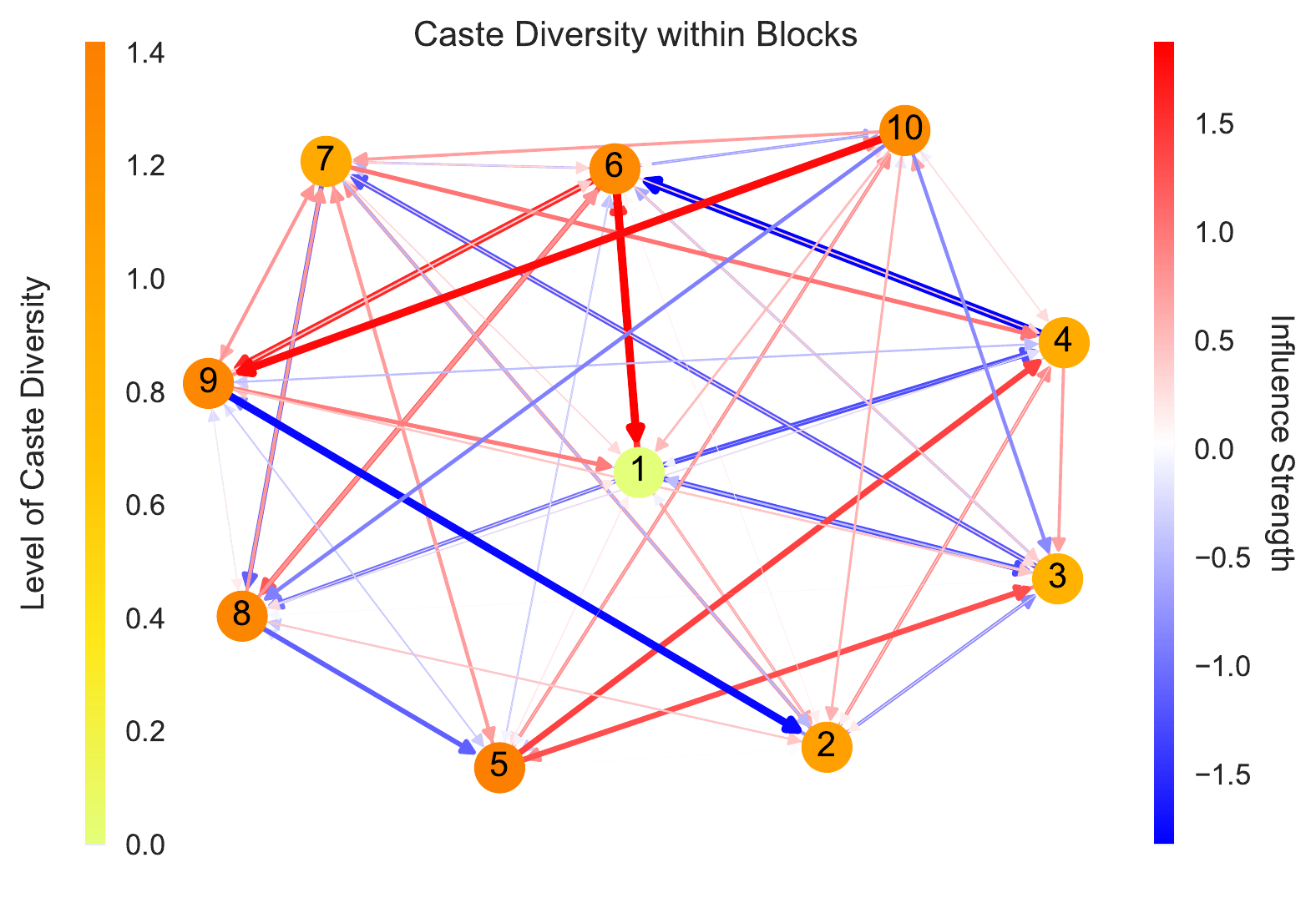} 
   \includegraphics[width=.32\linewidth]{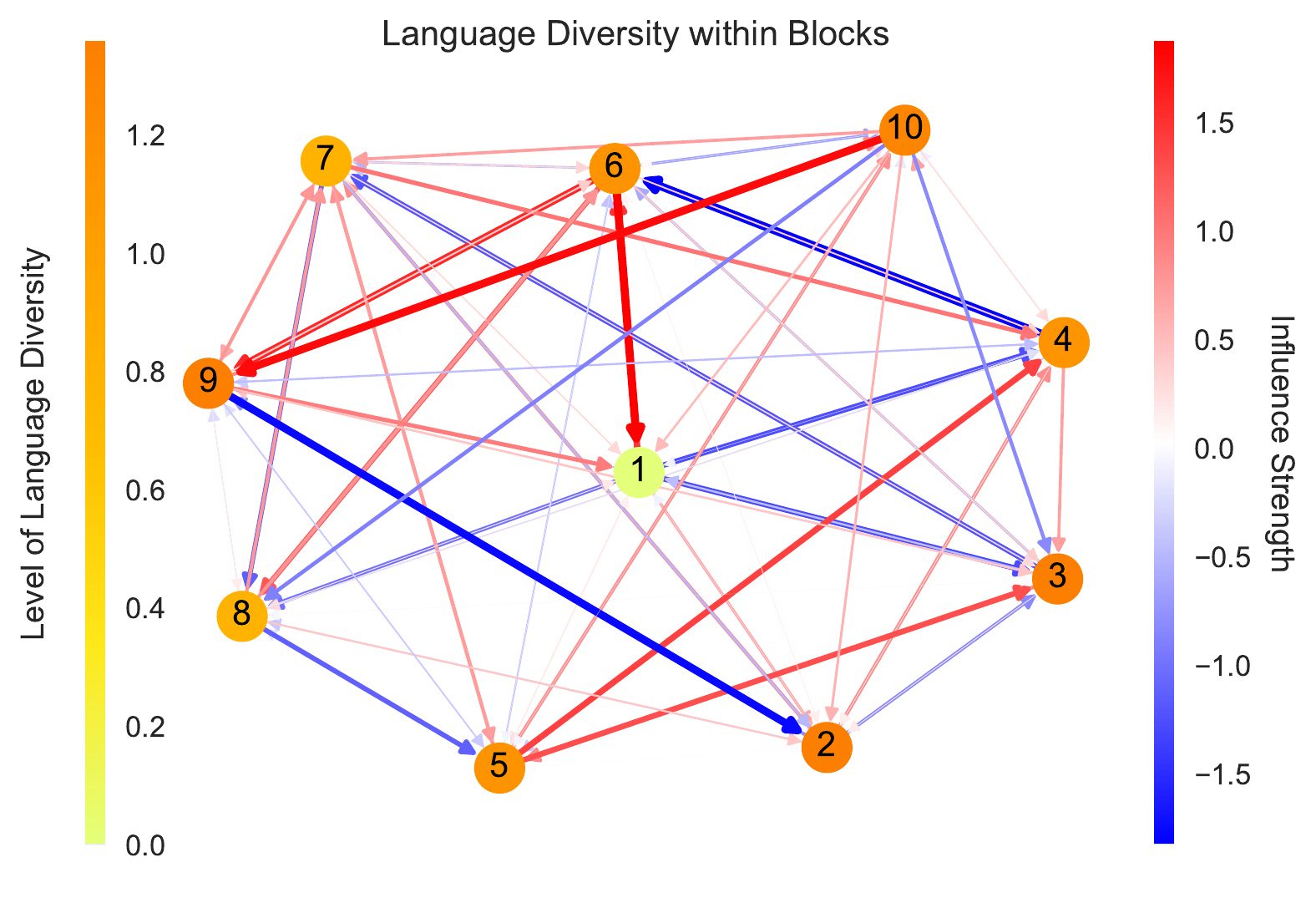} 
   \includegraphics[width=.32\linewidth]{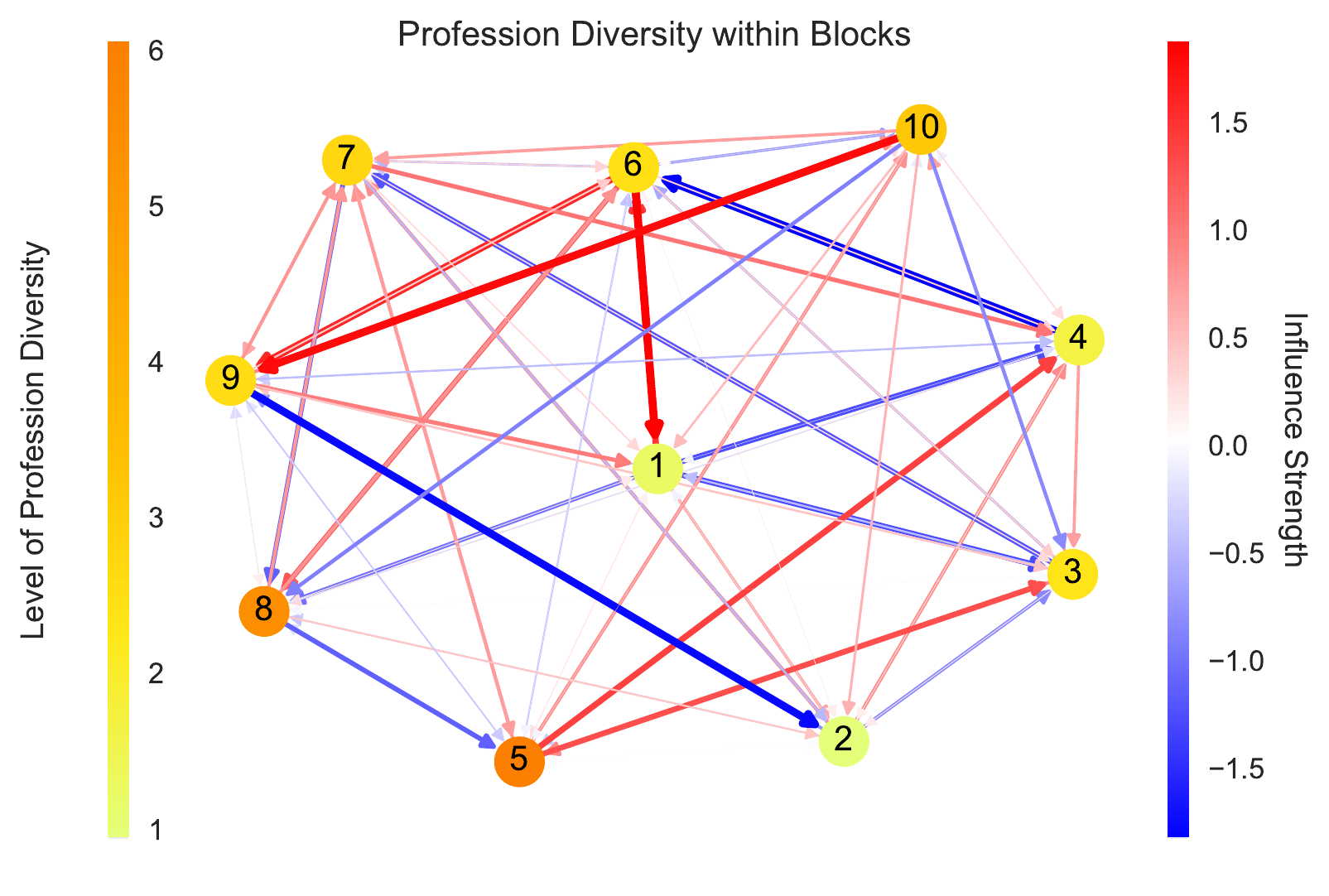} 
   \caption{Sociodemographic analysis of each social block and the social influence across social blocks. Each node represents a social block corresponding to the index shown in the previous in Table~\ref{blockChar}. The directed links represent the strength of social influence varying from strong negative (blue) to strong positive (red). The color of the node represents a measure of the sociodemographic characteristics within that social block. We display a subset of characteristics, including median age, gender ratio, caste diversity, language diversity, and profession diversity within each block. }
\label{netowrks}
\end{figure*}

We also use diversity and gender ratio to evaluate block characteristics for a specific example in Table~\ref{blockChar} and Figure~\ref{netowrks}, in addition to being used to evaluate the group attributes that are associated with different types of influence in Table~\ref{groupAtt}. More analysis in Figure~\ref{netowrks} is covered in the following section. 

We use normalized entropy to measure the diversity of different attributes. Normalized entropy is a metric used to capture the number of types of characteristics within each category while accounting for the frequency of each entity type within a category. It can be formulated by, $Q = - \frac{\sum_{i=1}^{q} p_{i} \log(p_{i})}{\sum_{i=1}^{q} \frac{1}{n_{i}} \log(\frac{1}{n_{i}})}$, where $q$ refers to the number of types within a category, $p_{i}$ refers to the probability of each type $i$, $f$ refers to the number of occurrences of each type $n_{i}$. 

The gender ratio ($R$) is measured within a block and is formulated by $R = \frac{r_{m}}{r_{f}}$, where $r_{m}$ and $r_{f}$ refer to the number of occurrences of males and females respectively. Thus, since $R$ is the ratio of males to females in a block, both a high or low gender ratio correspond to a high gender imbalance. 




\begin{figure}[h!t]
\centering
\subfloat[Influence matrix. The x-axis and y-axis correspond to the community index. The values are the strength of influence. The scale from blue to red corresponds with negative to positive influence. Darker colors correspond to a stronger influence. I f the y-axis is $\mathcal{V}_k$ and the x-axis is $\mathcal{V}_l$, the value corresponds to the influence from group $\mathcal{V}_k$ to $\mathcal{V}_l$.]{\includegraphics[width=0.6\linewidth]{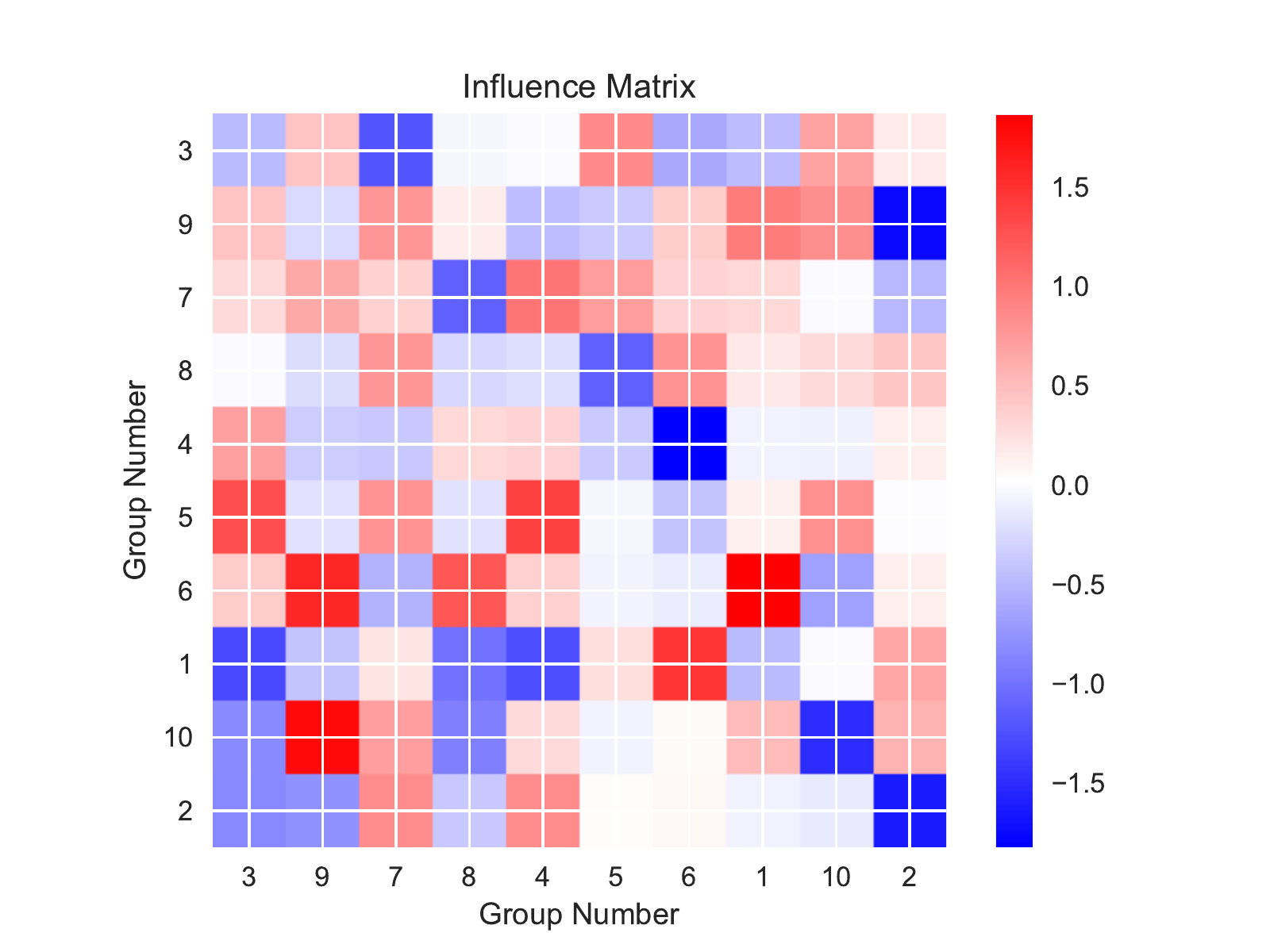}\label{fig:influence}} \\
\subfloat[Net influence into and out of each block. The x-axis and y-axis correspond to the block index and the aggregate strength of influence, respectively. 
The red and blue bars correspond to the in-flow and out-flow of social influence. ]{\includegraphics[width=0.6\linewidth]{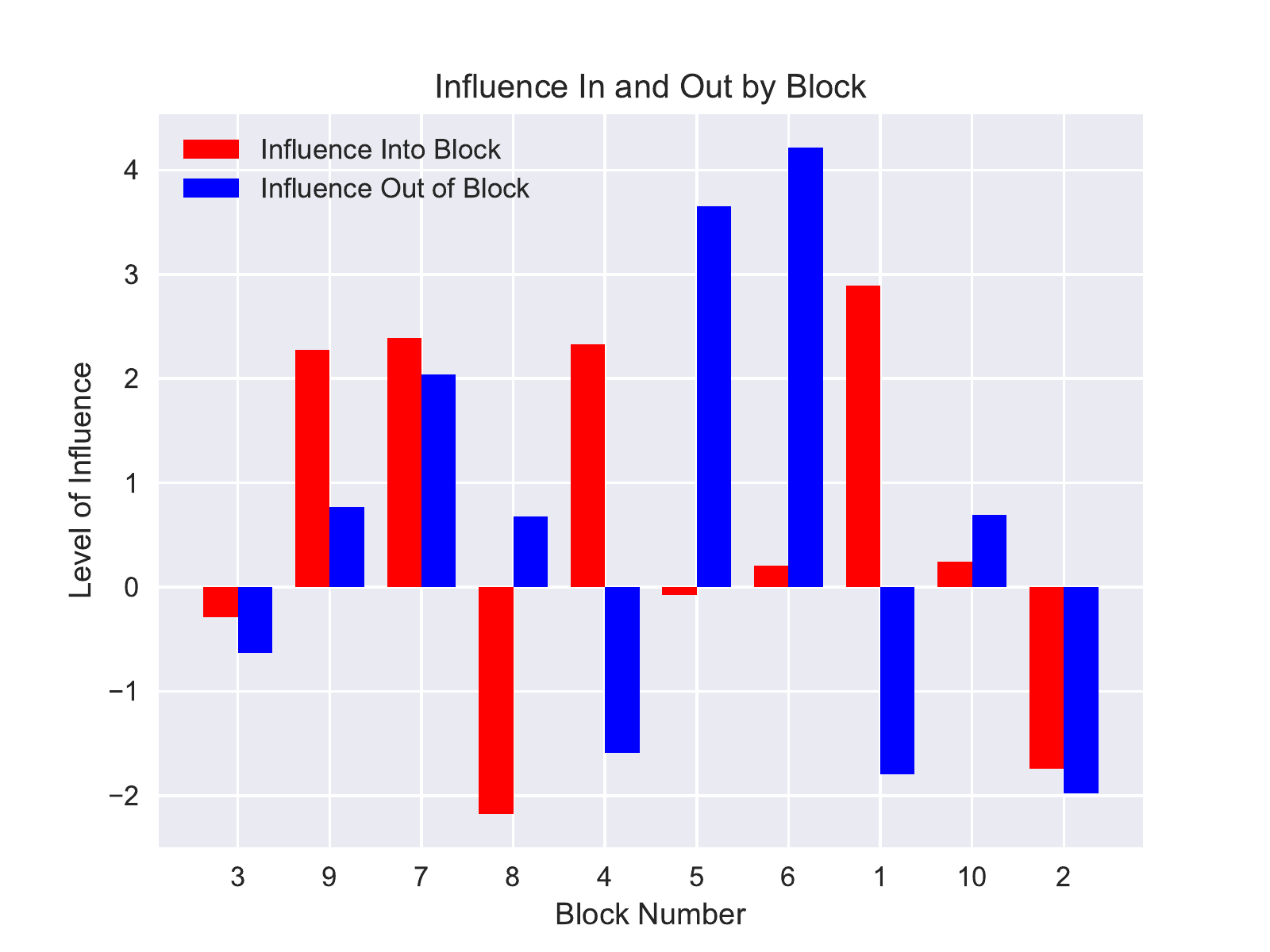}\label{fig:inout}}
\caption{Interaction matrix and influence matrix.}
\end{figure}

\paragraph*{Influence matrix and attributes}
The block-to-block influence, sorted by increasing block size, is displayed in Figure \ref{fig:influence}, where the strength of social influence, allowed to be either positive or negative, is shown. We can see some blocks influence other blocks ranging from strong negative influence to no influence, and to strong positive influence.

The total influence into and out of each block is depicted in Figure \ref{fig:inout}, which allows us to evaluate the aggregated influence a block receives and spreads (net positive, negative, or neutral). For example, we can see diverse, low-SES block five and senior, low-class block six with high output levels of positive influence, and diverse, middle-SES block eight receives a net high level of negative influence. We observe that some blocks have a stronger outgoing influence than other blocks and can perceive these as positive and negative influence leaders. Similar reasoning applies to characterize blocks that receive a high level of influence as follower blocks, furthermore observing the difference in net incoming and outgoing influence within each block as relating to its role in the block-to-block network. We refer to this to interpret different dynamics between social blocks, in addition to then pairing this information with demographic information to make further evaluations about block characteristics associated with different types of influence.

In Figure \ref{netowrks}, a subset of the sociodemographic features are displayed for each block, where the network of blocks is connected with varying degrees of influence between them. For example, we can see that lower median-age block four negatively influences the older median-age block six. The equal gender ratio block ten positively influences the similarly equal gender ratio block nine. Block ten influences block nine, where both blocks have similarly high caste diversity. Highly language diverse block six positively influences low language diverse block one. Lower professionally diverse block one negatively influences higher professionally diverse block three. 

\begin{table*}[h]
\small
\centering
  \caption{ Block characteristics example. SES is an abbreviation for socioeconomics status. The majority refers to the largest subset. Disadvantaged caste refers to lower castes, including the castes OBC (Other Backward Class) and Scheduled. Higher education refers to having education levels at PUC (pre-university course) and having a ``degree or above'' designation. Moderate and lower education levels include all levels below this, where moderate levels have more SSLC (Secondary School Leaving Certificate) levels, and PUC levels and lower levels have mostly primary school education levels.}
\label{blockChar}
{%
\begin{tabular}{|l|l|l|}
\hline
\textbf{Block}        & \textbf{Block Type}                      & \textbf{Attributes}                        \\ \hline
\multirow{2}{*}{1} & \multirow{2}{*}{Homogeneous, low-SES}       & only one disadvantaged caste and one language spoken   \\ \cline{3-3} 
          &                           & low profession diversity and education levels      \\ \hline
\multirow{2}{*}{2} & \multirow{2}{*}{Diverse, skilled, highly-educated}  & several different castes from many levels        \\ \cline{3-3} 
          &                           & diverse languages and diverse, high-skilled professions \\ \hline
\multirow{2}{*}{3} & \multirow{2}{*}{Senior, low-SES}          & majority disadvantaged caste               \\ \cline{3-3} 
          &                           & majority low skill-level professions in agriculture   \\ \hline
\multirow{3}{*}{4} & \multirow{3}{*}{Young, low-SES}          & younger average age, gender imbalanced block       \\ \cline{3-3} 
          &                           & majority lowest caste members, mostly natives      \\ \cline{3-3} 
          &                           & higher education                     \\ \hline
\multirow{3}{*}{5} & \multirow{3}{*}{Diverse, low-SES}         & diverse number of disadvantaged castes          \\ \cline{3-3} 
          &                           & moderate language diversity, moderate education     \\ \cline{3-3} 
          &                           & majority of jobs in agriculture                 \\ \hline
\multirow{3}{*}{6} & \multirow{3}{*}{Senior, low-SES}          & older average age, diverse in low castes         \\ \cline{3-3} 
          &                           & two languages spoken, very low education         \\ \cline{3-3} 
          &                           & lower-skilled professions                \\ \hline
\multirow{3}{*}{7} & \multirow{3}{*}{Homogeneous, low-SES}       & gender imbalanced, mostly disadvantaged caste      \\ \cline{3-3} 
          &                           & one language majority                  \\ \cline{3-3} 
          &                           & majority professions in agriculture and sericulture   \\ \hline
\multirow{3}{*}{8} & \multirow{3}{*}{Diverse, middle-SES}        & mostly one language                   \\ \cline{3-3} 
          &                           & caste diverse but mostly lower castes          \\ \cline{3-3} 
          &                           & diverse professions                   \\ \hline
\multirow{3}{*}{9} & \multirow{3}{*}{Diverse, highly-educated, low-SES} & disadvantaged caste majority               \\ \cline{3-3} 
          &                           & diverse jobs, higher-SES professions (teacher, priest) \\ \cline{3-3} 
          &                           & high education level, diverse languages         \\ \hline
\multirow{3}{*}{10} & \multirow{3}{*}{Homogeneous, low-SES}       & gender-balanced                       \\ \cline{3-3} 
          &                           & majority disadvantaged caste, only one language spoken  \\ \cline{3-3} 
          &                           & majority of professions in agriculture and sericulture  \\ \hline
\end{tabular}
}
\end{table*}

\begin{table*}[!htbp]
\small
  \caption{ Block attributes associated with different types of influence. Positive and negative influence refers to the type of influence from one block to another block. Self-influence refers to positive influence within the same block. Overlap refers to overlapping categories, such as caste type, profession type, education levels, or languages spoken. 
  }
  \begin{tabular}{p{1.5cm}|p{3.5cm}|p{3.5cm}|p{3.5cm}} \toprule
  \textbf{Attribute} & \textbf{Positive influence} & \textbf{Negative influence} & \textbf{Positive self-influence}
  \tabularnewline 
  \midrule
  \textbf{Gender} & similar gender distribution
  & gender-imbalanced block is more open to negative influence from gender-balanced block & large gender imbalance 
  \tabularnewline[1ex] \hline
  \textbf{Caste} & overlapping majority castes & lack of overlap in caste composition & majority village natives \tabularnewline [1ex]\hline
  \textbf{Profession} & profession overlap, in specialty jobs specifically; large professions diversity & professionally diverse block receives negative influence from a less professionally diverse block; lack of professional overlap causes a negative influence & high job diversity and higher-skilled jobs 
  \tabularnewline[1ex]\hline
  \textbf{Education} & large overlap in higher education level & higher educated block receives negative influence from less educated block & higher education level
 \tabularnewline[1ex]\hline
  \textbf{Language} & overlapping language & lack of overlap in language & language diversity
  \tabularnewline[1ex]\hline
  \textbf{Age} & none & older-age block can receive negative influence from younger-age block & younger age 
  \tabularnewline[1ex]
  \bottomrule
  \end{tabular}
  \label{groupAtt}
\end{table*}

By analyzing several examples in this manner using block characteristic composition and observing the types and patterns of influence, several general trends arise, as depicted in Table \ref{groupAtt}. The block attributes most frequently associated with different types of influence are summarized into key trends. Positive influence occurs when two blocks overlap in the following characteristics: gender distribution, majority castes, professions, high profession diversity, highly educated, highly-skilled jobs, and mother tongue languages. Negative influence frequently occurs when two blocks have a lack of overlap in the following characteristics: gender distribution, caste composition, profession diversity level, education levels, and average age. Furthermore, the direction of negative influence is most frequently observed from a low-SES block to a high-SES block. Additionally, we frequently observe positive self-influence, which is from a block to itself, and this occurs when a block is characterized by a younger average age, highly-educated, high job diversity, higher-skilled jobs, high language diversity, large gender imbalance, and having a large number of village natives. 

 These trends, when paired with block type characterizations, lead to interesting associations, such as block-to-block perceptions of lower or higher SES groups with influence. Blocks of the higher SES group designation more frequently received negative influence from lower-SES blocks. Blocks of similar SES, especially higher SES, had a more frequent positive influence between them. High-SES blocks also had more frequent positive self-influence. 

{These findings suggest some marketing strategies that take into account the underlying communities. For example, the microfinance institution could organize separate information sessions for the high-SES and low-SES groups to take advantage of the positive influence between groups that share similar characteristics, while avoiding the negative influence that occurs across the different communities. Moreover, if the microfinance institution is to introduce the product into other villages (as a new product), they should send the information to individuals with the following characteristics: (1) high-SES with less low-SES neighbors, (2) individuals who speak a diverse set of languages, and (3) communities with similar gender ratios.}

\section{Applications and future works}
\label{sec:conclusion}
{ Role theory postulates that the interactions of individuals depend on their roles and behaviors of interest. To conceptualize this idea, we use the underlying community structures to capture the ``roles'', which affect the particular decision-making processes of individuals. }Specifically, we develop the Stochastic Block Influence Model, which infers two types of hidden relationships: (1) block-to-block interaction, and (2) block-to-block influence on decision-making. Moreover, our model flexibly allows for both positive and negative social influence. The latter is more common in practice but has been ignored by the contagion models in the literature \cite{centola2007complex,kempe2003maximizing,banerjee2013diffusion}. In the adoption of microfinance examples we present, the inferred block-to-block influence offers insights into how different social blocks exert influence on individuals' decision-making. The framework has far-reaching practical impacts for understanding patterns of influence across communities and identifying the crucial characteristics of influential individuals for several applications. {To name a few: 
\begin{enumerate}
  \item Practitioners and researchers can identify the most influential communities (e.g., leaders and followers) and understand the dynamics among different communities that are not available {nor observable} without our model. 
  \item Marketing campaigner can investigate in which sociodemographics predict positive or negative social influence, and utilize this information when introducing the product to a new market. 
  \item Marketing firms can use the influence of each individual to decide whom to target for campaigns \cite{leng2018contextual}. For example, in marketing campaigns, we should advertise to individuals who spread positive aggregate influence. 
  \item For policy-makers, the behavioral model in our paper can be used to perform counterfactual predictions for network interventions to predict responses to new policies.  
\end{enumerate}
}

Our method is not without limitations and hence opens up several directions for future studies. First, future research can easily adapt SBIM to accommodate a more complicated stochastic block model, such as a degree-corrected SBM or a power-law regularized SBM. Second, a scalable inference method as an alternative to NUTS sampling will help to improve the efficiency and scalability of SBIM. Third, future research can extend SBIM to a dynamic model, where the influence matrix varies with time and distances from the source of information. Lastly, for computer scientists and social scientists who have access to similar types of data, but in different settings (e.g., different behaviors and collected in different countries), it will be interesting to apply and compare the influence matrices to see if there exists any generalizable pattern to support existing contagion and decision-making theories.

\bibliographystyle{ACM-Reference-Format}
\bibliography{main} 

\end{document}